\documentclass[11pt,a4paper]{article}
\pdfoutput=1 

\RequirePackage{ifpdf} 
\usepackage{amsmath} 
\usepackage{mathtools}
\usepackage{soul}
\usepackage{jheppub}
\usepackage{pstricks}
\usepackage[final]{pdfpages} 
\usepackage{ifpdf} 
\usepackage{slashed}

\usepackage{color,xcolor}
\definecolor{urlblue}{rgb}{0.2,0.4,0.7}
\definecolor{citegreen}{rgb}{0,0.6,0.2}
\definecolor{linkred}{rgb}{0.9,0.2,0.1}
\usepackage{hyperref}
\hypersetup{
colorlinks=true, citecolor=blue, linkcolor=blue, urlcolor=urlblue}

\usepackage{graphics}
\usepackage{etoolbox} 
\usepackage{fixmath}
\usepackage{psfrag}

\usepackage{notoccite} 

\usepackage{amsfonts}
\usepackage{autobreak}
\usepackage{caption}
\usepackage{subcaption}

\usepackage{tikz}
\usetikzlibrary{positioning,arrows}
\usetikzlibrary{decorations.pathmorphing}
\usetikzlibrary{decorations.markings}
\usetikzlibrary{shapes.geometric}
\tikzset{
    vector/.style={decorate, decoration={snake}, draw},
    provector/.style={decorate, decoration={snake,amplitude=2.5pt}, draw},
    antivector/.style={decorate, decoration={snake,amplitude=-2.5pt}, draw},
    fermion/.style={draw=black, postaction={decorate},decoration={markings,mark=at position .55 with {\arrow[draw=black]{>}}}},
    fermionbar/.style={draw=black, postaction={decorate},
                       decoration={markings,mark=at position .55 with {\arrow[draw=black]{<}}}},
    fermionnoarrow/.style={draw=black},
    gluon/.style={decorate, draw=black,decoration={coil,amplitude=4pt, segment length=5pt}},
    scalar/.style={dashed,draw=black, postaction={decorate},decoration={markings,mark=at position .55 with {\arrow[draw=black]{>}}}},
    scalarbar/.style={dashed,draw=black, postaction={decorate},decoration={markings,mark=at position .55 with {\arrow[draw=black]{<}}}},
    scalarnoarrow/.style={dashed,draw=black},
    electron/.style={draw=black, postaction={decorate},decoration={markings,mark=at position .55 with {\arrow[draw=black]{>}}}},
    bigvector/.style={decorate, decoration={snake,amplitude=4pt}, draw},
}

\title{Two-loop amplitudes for di-Higgs and di-pseudo-Higgs productions through quark annihilation in QCD}

\author{Taushif Ahmed$^{a,1}$, V. Ravindran$^{b,2}$, Aparna Sankar$^{b,3}$ and Surabhi Tiwari$^{b,4}$}
\emailAdd{$^1$taushif.ahmed@unito.it, $^2$ravindra@imsc.res.in, $^3$aparnas@imsc.res.in, $^4$surabhit@imsc.res.in}

\affiliation{$^a$Dipartimento di Fisica and Arnold-Regge Center, Universit\`a di Torino, 
\\ and INFN, Sezione di Torino, Via Pietro Giuria 1, I-10125 Torino, Italy \\ 
$^b$The Institute of Mathematical Sciences, HBNI, IV Cross Road, Taramani, Chennai 600113, India}

\preprint{IMSc/2021/10/08}

\abstract{Through this article, we present the two-loop massless QCD corrections to the production of di-Higgs and di-pseudo-Higgs boson through quark annihilation in the large top quark mass limit. Within dimensional regularisation, we employ the non-anticommuting $\gamma_5$ and treat it under the Larin prescription. We discover the absence of any additional renormalisation, so-called contact renormalisation, that could arise from the short distance behaviour of two local operators. This finding is in corroboration with the operator product expansion. By examining the results, we discover the lack of similarity in the highest transcendentality weight terms between these finite remainders and that of a pair of half-BPS primary operators in maximally supersymmetric Yang-Mills theory. We need these newly computed finite remainders to calculate observables involving di-Higgs or di-pseudo-Higgs at the next-to-next-to-leading order. We implement the results to a numerical code for further phenomenological studies.}

\begin{document}
\allowdisplaybreaks[4]
\unitlength1cm
\keywords{Higgs boson, pseudo-scalar, two-loop, QCD, HEFT}
\maketitle
\flushbottom


\def\D{{\cal D}}
\def\DD{\overline{\cal D}}
\def\g{\overline{\cal G}}
\def\gm{\gamma}
\def\M{{\cal M}}
\def\ep{\epsilon}
\def\epm1{\frac{1}{\epsilon}}
\def\epm2{\frac{1}{\epsilon^{2}}}
\def\epm3{\frac{1}{\epsilon^{3}}}
\def\epm4{\frac{1}{\epsilon^{4}}}
\def\unM{\hat{\cal M}}
\def\ashat{\hat{a}_{s}}
\def\asmur{a_{s}^{2}(\mu_{R}^{2})}
\def\sigbar{{{\overline {\sigma}}}\left(a_{s}(\mu_{R}^{2}), L\left(\mu_{R}^{2}, m_{H}^{2}\right)\right)}
\def\sigbarn{{{{\overline \sigma}}_{n}\left(a_{s}(\mu_{R}^{2}) L\left(\mu_{R}^{2}, m_{H}^{2}\right)\right)}}
\def\unas{ \left( \frac{\hat{a}_s}{\mu_0^{\epsilon}} S_{\epsilon} \right) }
\def\rnM{{\cal M}}
\def\bt{\beta}
\def\cD{{\cal D}}
\def\cC{{\cal C}}
\def\ca{\text{\tiny C}_\text{\tiny A}}
\def\cf{\text{\tiny C}_\text{\tiny F}}
\def\ct{{\red []}}
\def\sv{\text{SV}}
\def\murOmu{\left( \frac{\mu_{R}^{2}}{\mu^{2}} \right)}
\def\bb{b{\bar{b}}}
\def\bt0{\beta_{0}}
\def\bt1{\beta_{1}}
\def\bt2{\beta_{2}}
\def\bt3{\beta_{3}}
\def\gm0{\gamma_{0}}
\def\gm1{\gamma_{1}}
\def\gm2{\gamma_{2}}
\def\gm3{\gamma_{3}}
\def\nn{\nonumber}
\def\l{\left}
\def\r{\right}
\def\T{{\cal Z}}    
\def\U{{\cal Y}}
\def\qgraf{{\fontfamily{qcr}\selectfont
QGRAF}}
\def\python{{\fontfamily{qcr}\selectfont
PYTHON}}
\def\form{{\fontfamily{qcr}\selectfont
FORM}}
\def\reduze{{\fontfamily{qcr}\selectfont
REDUZE2}}
\def\kira{{\fontfamily{qcr}\selectfont
Kira}}
\def\litered{{\fontfamily{qcr}\selectfont
LiteRed}}
\def\fire{{\fontfamily{qcr}\selectfont
FIRE5}}
\def\air{{\fontfamily{qcr}\selectfont
AIR}}
\def\mint{{\fontfamily{qcr}\selectfont
Mint}}
\def\hepforge{{\fontfamily{qcr}\selectfont
HepForge}}
\def\arXiv{{\fontfamily{qcr}\selectfont
arXiv}}
\def\Python{{\fontfamily{qcr}\selectfont
Python}}
\def\anci{{\fontfamily{qcr}\selectfont
Finite\_ppbk.m}}
\def\finiteflow{{\fontfamily{qcr}\selectfont
FiniteFlow}}

\newcommand{\dis}{}
\newcommand{\overbar}[1]{mkern-1.5mu\overline{\mkern-1.5mu#1\mkern-1.5mu}\mkern
1.5mu}


\section{Introduction}
\label{sec:intro}
The detailed exploration of the discovered scalar resonance~\cite{ATLAS:2012yve,CMS:2012qbp} with a mass of $125.09 \pm 0.24$ GeV~\cite{ATLAS:2016neq} at the Large Hadron Collider (LHC) is underway with spectacular success. After establishing the coupling of the Higgs boson to all electroweak gauge bosons~\cite{ATLAS:2016neq}, top~\cite{CMS:2018uxb,ATLAS:2018mme} and bottom quarks~\cite{ATLAS:2018kot,CMS:2018nsn}, tau lepton~\cite{CMS:2017zyp,ATLAS:2018ynr}, now LHC has its focus on the measurement of its coupling with muon~\cite{CMS:2020xwi,ATLAS:2020fzp}. Reconstructing the Higgs potential experimentally and consequently pinning down the mechanism of electroweak symmetry breaking is one of the most important objectives in current times. This requires the precise determination of Higgs self-coupling which can be accessed directly through the production of Higgs boson pair. While the determination of quartic self-coupling is difficult at the LHC, the probe of trilinear coupling is certainly possible. Through the current run~\cite{CMS:2018ipl,ATLAS:2019qdc,CMS:2020tkr}, the limits on the latter is improved significantly. The theoretical value of $\lambda$, which dictates the self-coupling, can be determined from the available precise results of the mass and vacuum expectation value of the scalar Higgs boson. However, it is important to verify it independently through experiment.

In the Standard Model (SM), the Higgs boson pair is dominantly produced through gluon fusion containing a top quark loop. The production cross-section at leading order (LO) in QCD has been computed exactly in refs.~\cite{Eboli:1987dy,Glover:1987nx,Plehn:1996wb,Djouadi:1999rca}. Due to the presence of massive top quark loop, computing it exactly beyond the LO is a nontrivial task. To reduce the complexity while capturing the dominant contribution, the Higgs Effective Field Theory (HEFT) has been designed by integrating out the top quark loop by treating its mass as infinitely large. In ref.~\cite{Dawson:1998py}, the next-to-LO (NLO) corrections in HEFT were first computed by keeping the LO exact. Later, almost after two decades, the NLO corrections by keeping the exact top quark mass were achieved in refs.~\cite{Borowka:2016ehy,Borowka:2016ypz,Baglio:2018lrj,Baglio:2020ini}. In refs.~\cite{Heinrich:2017kxx,Jones:2017giv,Heinrich:2019bkc,Heinrich:2020ckp}, the parton shower was also included to make the prediction more robust. The threshold resummation effect was investigated in ref.~\cite{Shao:2013bz}.

In HEFT, the NNLO QCD corrections were computed in refs.~\cite{deFlorian:2013uza,deFlorian:2013jea,Grigo:2014jma,Grigo:2015dia,deFlorian:2016uhr}. The relevant Wilson coefficients were presented in ref.~\cite{Grigo:2014jma}. By keeping the exact top quark dependence up to NLO and in the double real radiation, the NNLO correction was further improved in ref.~\cite{Grazzini:2018bsd} which made use of the result from ref.~\cite{deFlorian:2016uhr}. These results were further improved by including the soft-gluon resummed contribution in ref.~\cite{deFlorian:2018tah}. The differential results at NNLO can be found in refs.~\cite{Li:2013flc,Maierhofer:2013sha,deFlorian:2015moa}. Quite recently, the focus has moved towards the N$^3$LO corrections within HEFT. The computation of the two-loop virtual correction~\cite{Banerjee:2018lfq} and the four loop matching coefficient for the effective coupling of two Higgs bosons and gluons~\cite{Spira:2016zna,Gerlach:2018hen} has been achieved. Within HEFT, the N$^3$LO results were computed in refs.~\cite{Chen:2019lzz,Chen:2019fhs}, where the result of ref.~\cite{Chen:2019fhs} is NLO-improved employing the results of refs.~\cite{Heinrich:2017kxx,Heinrich:2019bkc}. Recently, in ref.~\cite{Davies:2021kex}, the NNLO cross-section has been improved by computing the higher order terms in top mass expansion.

Although the gluon initiated process for di-Higgs production is dominant, there is a need of incorporating the subdominant channels to make the theoretical predictions as precise as possible. The first step towards this direction was successfully attempted in ref.~\cite{Ajjath:2018ifl}, where the NNLO correction to the process initiated by the bottom quark annihilation was computed. The Higgs boson was assumed to couple to the bottom quark through non-zero Yukawa coupling, otherwise the latter was treated as a massless entity. In this article, we address the computation of two-loop QCD correction to di-Higgs production through quark annihilation in HEFT. On top of its phenomenological relevance, we intend to discover whether there is a need for any additional renormalisation arising from the short distance behaviour of two composite operators. For instance, for the gluon initiated process, its necessity was discovered in refs.~\cite{Zoller:2016iam,Banerjee:2018lfq}. In the former reference, it was found out from the operator product expansion (OPE). In this article, we want to find whether there is a need for any additional contact renormalisation arising from short distance behaviour for the quark initiated channel. Moreover, we intend to determine whether any subtle issue shows up owing to working in an effective theory, an example of this can be found out from ref.~\cite{Ahmed:2020kme} where one of the authors demonstrated the need for an additional four-point effective vertex.
\\~

In addition to the scalar Higgs, the exploration of the Higgs sector at the LHC includes the search for a CP-odd Higgs boson which is often called a pseudo-scalar. In certain beyond the SM (BSM) scenarios, for instance the minimally supersymmetric SM (MSSM), the Higgs sector contains a pseudo-scalar in addition to two neutral and two charged scalars. It is important to have theoretical predictions associated with a CP-odd Higgs boson as precise as with the CP-even counterpart. While the observables associated with the latter have been computed to a very high accuracy, the progress involving di-pseudo-scalar is not up to the requirement. One of the reasons is the handling of chiral quantity in dimensional regularisation where space-time dimension is analytically continued to $d=4-2\epsilon$ with $\epsilon$ being a complex number. In ref.~\cite{Ahmed:2019udm}, some of us demonstrated how to simplify the calculation of radiative corrections involving a chiral quantity. For a single CP-odd Higgs boson, the production cross-section is available to N$^3$LO under soft-virtual approximation~\cite{Harlander:2002vv,Anastasiou:2002wq,Ravindran:2003um,Ahmed:2015qpa,Ahmed:2016otz,Ahmed:2015qda}. For the di-pseudo-scalar, the NLO in heavy top limit was obtained in ref.~\cite{Dawson:1998py}. The first step to go beyond the NLO was attempted in ref.~\cite{Bhattacharya:2019oun} where the two-loop QCD correction for the production of di-pseudo-scalar through gluon fusion was achieved. In this article, we address the computation in the quark annihilation channel in the heavy top limit up to two loops. In the process, we discover the absence of any contact renormalisation, consistent with the analysis through OPE in ref.~\cite{Zoller:2013ixa}.

In section~\ref{sec:kin}, we discuss the kinematics of the four-point amplitudes and the dimensionless variables which are introduced for future reference. We discuss the effective Lagrangian in the heavy top limit in section~\ref{sec:th}. A short discussion on handling $\gamma_5$ is also included. In section~\ref{sec:psps}, we introduce the form factor decomposition and consequently the projector that is employed to get the form factors up to two-loop in QCD. We briefly discuss the method of computing the loop amplitude that is adopted in this article. The ultraviolet renormalisation is performed in section~\ref{sec:op-ren}. We discuss the infrared factorisation in terms of universal subtraction operator and some properties of the ultraviolet renormalised form factors in section~\ref{sec:IR}. Finally we perform the numerical analysis of the results in section~\ref{sec:Num}. Then we summarise and draw the conclusion in section~\ref{sec:concl}.

\section{Preliminaries}
\label{sec:kin}
We consider the production of di-scalar ($\phi$) and di-pseudo-scalar ($\tilde\phi$) Higgs boson through quark annihilation
\begin{align}
\label{eq:process}
&q(p_1)+\bar q(p_2) \rightarrow \phi(p_3) + \phi(p_4)\nonumber\\
&q(p_1)+\bar q(p_2) \rightarrow \tilde\phi(p_3) + \tilde\phi(p_4)\,,
\end{align}
where we denote the four-momentum by $p_i$ satisfying the on-shell conditions $p_1^2=p_2^2=0$ and $p_3^2=p_4^2=m^2$. We represent quark (anti-quark) by $q$($\bar q$). The parameter $m$ denotes the mass of the scalar or pseudo-scalar depending on the process. We define the Mandelstam variables through
\begin{align}
\label{eq:mandelstam}
s \equiv (p_1+p_2)^2,\quad t \equiv (p_1-p_3)^2, \quad u \equiv (p_2-p_3)^2
\end{align}
which, owing to the momentum conservation, satisfy
\begin{align}
\label{eq:momconserv}
s+t+u=2m^2\,.
\end{align}
For future requirements, we introduce three dimensionless variables through
\begin{align}
\label{eq:xyz}
s = m^2 \frac{(1+x)^2}{x},\quad t=-m^2 y,\quad u=-m^2 z\,.
\end{align}
These partonic channels contribute to the production of di-scalar and di-pseudo-scalar Higgs boson at a hadronic collision. In this article, we calculate the scattering amplitudes within the framework of the heavy-top effective theory which we now turn into.

\section{Theoretical framework under heavy top limit}
\label{sec:th}
In the heavy-top effective field theory, that is obtained by integrating out the top quark appearing in the loop, the interacting Lagrangian density describing the one and two scalar Higgs bosons is given by
\begin{align}
\label{eq:effLags}
{\cal L}^{\phi}_{\rm eff} = -\frac{1}{4} G^{\mu\nu}_a G_{a,\mu\nu} \left( \frac{1}{v}C_H \phi (x) - \frac{1}{v^2} C_{HH} \phi^2(x) \right)\,.
\end{align}
The quantity $G^{\mu\nu}_a$ denotes the gluon field strength tensor and $v=246$ GeV is the vacuum expectation value of the scalar Higgs field. In moving from exact to the effective theory where we work in $n_f=5$ light flavor, all the top quark mass dependence is encapsulated within the Wilson coefficients $C_H$ and $C_{HH}$. These are determined by matching the effective $n_f=5$ and full $(n_f+1)$ flavor theories. Furthermore, this matching can be done order by order in perturbation theory which in powers of renormalised strong coupling constant, $a_s(\mu_R^2) = \alpha_s(\mu_R^2)/(4\pi)$, are found to be~\cite{Grigo:2014jma,Djouadi:1991tka,
Kramer:1996iq,Chetyrkin:1997iv,Spira:2016zna,Gerlach:2018hen}
\begin{align}
\label{eq:CH}
C_H &= -\frac{4 a_s}{3}  \Bigg[ 1 + a_s   \Big\{ 5C_A-3C_F\Big\} + {\cal O}(a_s^2)\Bigg]\,.
\end{align}
In the aforementioned results, $m_t$ is the top quark mass in $\overline{\rm MS}$ scheme determined at the renormalisation scale $\mu_R$. We denote the quadratic Casimirs of SU($n_c$) group in the fundamental and adjoint representations by $C_A=n_c$ and $C_F=(n_c^2-1)/(2n_c)$, respectively. The constant $T_f$ equals $1/2$. We skip presenting the expression of $C_{HH}$ as it is not required in our current calculation, as discussed in the upcoming section.

In an analogous way, in this effective theory the pseudo-scalar Higgs boson ($\tilde\phi$) can also be described through a Lagrangian containing a gluonic ($O_G$) and a quarkonic ($O_J$) operator. The effective Lagrangian~\cite{Chetyrkin:1998mw} reads as
\begin{align} 
\label{eq:effLagps}
{\cal L}^{\tilde \phi}_{\rm eff} = \tilde\phi(x) \Big[ -\frac{1}{8}
  {C}_{G} O_{G}(x) - \frac{1}{2} {C}_{J} O_{J}(x)\Big]\,,
\end{align}
where the operators are defined in terms of gluon field strength tensor  and light quark wave function ($\psi$) as
\begin{align}
\label{eq:OGOJ}
  O_{G}(x) = G^{\mu\nu}_a \tilde{G}_{a,\mu
    \nu} \equiv  \epsilon_{\mu \nu \rho \sigma} G^{\mu\nu}_a G^{\rho
    \sigma}_a\, ,\qquad
  O_{J}(x) = \partial_{\mu} \left( \bar{\psi}
    \gamma^{\mu}\gamma_5 \psi \right)  \,.
\end{align}
The transition from exact to effective theory introduces two Wilson coefficients, $C_G$ and $C_J$ which can be expanded in powers of strong coupling constant:
\begin{align}
\label{eq:CGCJ}
  & C_{G} = -a_{s} 2^{\frac{5}{4}} G_{F}^{\frac{1}{2}} {\rm \cot} \beta
    \nonumber\\
  & C_{J} = - \left[ a_{s} C_{F} \left( \frac{3}{2} - 3\ln
    \frac{\mu_{R}^{2}}{m_{t}^{2}} \right) + a_s^2 C_J^{(2)} + \cdots \right] C_{G}\, .
\end{align}
The absence of higher-order terms beyond one-loop in $C_G$ is guaranteed by the Adler-Bardeen theorem~\cite{Adler:1969er}. We denote the Fermi constant by $G_F$ and mixing angle in a generic two-Higgs doublet model by $\beta$. The superscript of $C_J$ indicates the corresponding perturbative order. In the upcoming section, we discuss how we compute the di-scalar and di-pseudo-scalar amplitudes within this framework of effective theory.

The presence of the chiral quantity, Dirac's $\gamma_5$ (and Levi-Civita symbol $\epsilon_{\mu\nu\rho\sigma}$), which are inherently defined in 4-dimensions faces the issue of translating properly into $d$-dimensions in dimensional regularisation (DR)~\cite{tHooft:1972tcz,Bollini:1972ui}. One of the most popular ways of handling it to define it following the 't Hooft-Veltman~\cite{tHooft:1972tcz} and Breitenlohner-Maison~\cite{Breitenlohner:1977hr} scheme as
\begin{align}
\label{eq:gamma5-defn}
\gamma_5=\frac{i}{4!} \epsilon_{\mu\nu\rho\sigma} \gamma^\mu \gamma^\nu \gamma^\rho \gamma^\sigma\,.
\end{align}
This definition has many profound implications which we need to address. Firstly, it breaks the hermiticity property of the operator which is cured by replacing~\cite{Larin:1991tj,Akyeampong:1973xi}
\begin{align}
\label{eq:gmug5}
\gamma_\mu\gamma_5 = \frac{i}{3!} \epsilon_{\mu\nu\rho\sigma} \gamma^\nu\gamma^\rho\gamma^\sigma
\end{align}
instead of directly using the definition \eqref{eq:gamma5-defn}. The contraction of Levi-Civita symbols is done in $d$-dimensions using
\begin{align}
\label{eq:contract-levi}
\epsilon_{\mu_1\nu_1\rho_1\sigma_1} \epsilon^{\mu_2\nu_2\rho_2\sigma_2} = 4! \delta^{\mu_2}_{[\mu_1} \cdots \delta^{\sigma_2}_{\sigma_1]}
\end{align}
where the indices of the metric tensor on the right-hand side are treated in $d$ dimensions~\cite{Zijlstra:1992kj}. The anti-symmetrisation of the Lorentz indices is denoted through $[\, ]$.


\section{Four-point amplitudes}
\label{sec:psps}
Performing the Lorentz covariant decomposition, we can write down the general form of the amplitude for these scattering processes as
\begin{align}
\label{eq:FFdecom-scalar}
{\cal A}_{ij,h} =  \bar v(p_2) \Gamma_h u(p_1) \delta_{ij}\,, 
\end{align}
where $h$ can either be $\phi$ or $\tilde\phi$. By applying on-shell Dirac equation and keeping in mind the conservation of chirality along the massless quark line along with the parity even final states, it is straightforward to guess the form of $\Gamma_h$ as
\begin{align}
\label{eq:Gammah}
\Gamma_h = F_h \slashed{p}_3\,.
\end{align}
The form factor $F_h$ can be calculated order by order in perturbation theory.
At the leading order, both of these processes are one-loop induced in HEFT since these start at two-loop in exact theory. Our \textit{goal} is to compute the two-loop QCD contributions to $F_h$ in the HEFT. To compute the form factor (FF), we can apply an appropriate projector on ${\cal A}_{ij,h}$, which in this case comes out to be
\begin{align}
    \label{eq:projector}
    {\cal P} = \frac{1}{2(tu-m^4)} \bar u(p_1) \slashed{p}_3 v(p_2)\,.
\end{align}

We start the computation by generating the set of Feynman diagrams in heavy quark effective theory using \qgraf~\cite{Nogueira:1991ex} that provides the expression in a symbolic form. For the production of di-Higgs through quark annihilation, we categorise the diagrams into class-A and class-B depending on
\begin{figure}[h]
	\centering
	\begin{subfigure}[c]{.25\textwidth}
		\raisebox{-\height}{\includegraphics[width=0.8\textwidth]{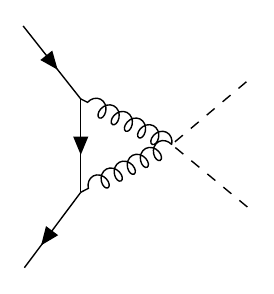}}
	\end{subfigure}
	\caption{Sample diagram for $q\bar q\rightarrow\phi\phi$ and $q\bar q\rightarrow\tilde\phi\tilde\phi$ at the LO. These vanish to all orders. The dotted lines either denote scalar Higgs or pseudo-scalar Higgs boson depending on the process under consideration.}
	\label{fig:qqhh-classA}
\end{figure}
whether those are proportional to $C_{HH}$ or $C_H^2$, respectively. The diagrams at the LO belong to class-A, as shown in figure~\ref{fig:qqhh-classA}.
\begin{figure}[h]
	\centering
	\begin{subfigure}[c]{.2\textwidth}
		\raisebox{-\height}{\includegraphics[width=0.8\textwidth
		]{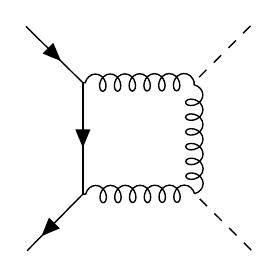}}
	\end{subfigure}
\hspace{1cm}
	\begin{subfigure}[c]{.2\textwidth}
		\raisebox{-\height}{\includegraphics[width=0.8\textwidth,angle=270]{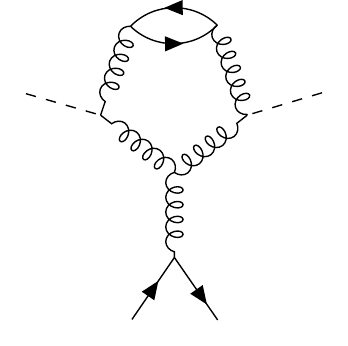}}
	\end{subfigure}
\hspace{1cm}
	\begin{subfigure}[c]{.2\textwidth}
		\raisebox{-\height}{\includegraphics[width=0.8\textwidth,angle=270]{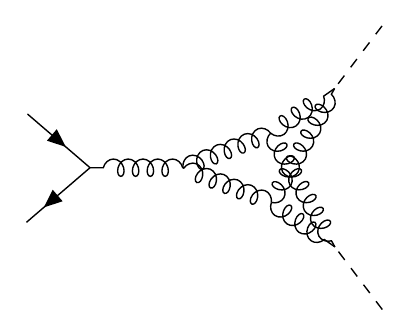}}
	\end{subfigure}
\hspace{1cm}
	\begin{subfigure}[c]{.2\textwidth}
		\raisebox{-\height}{\includegraphics[width=0.8\textwidth]{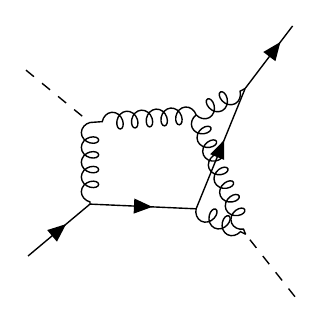}}
	\end{subfigure}
\hspace{1cm}
	\begin{subfigure}[c]{.2\textwidth}
		\raisebox{-\height}{\includegraphics[width=1.3\textwidth]{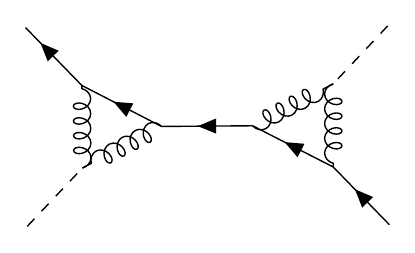}}
	\end{subfigure}
	\caption{Sample diagrams for $q\bar q\rightarrow\phi\phi$ at NLO (first diagram) and NNLO which belong to class-B and are proportional to $C_{H}^2$}
	\label{fig:qqhh-classB}
\end{figure}
The class of diagrams belonging to this category vanish to all orders since this part of the amplitude is proportional to the mass of external quarks which is zero in massless QCD and hence only the class-B diagrams contribute. At NLO, there are 5 diagrams, and at NNLO there are 143. We have shown some sample diagrams of this category in figure~\ref{fig:qqhh-classB}. Before moving to pseudo-scalar, let us mention that we do not need to compute the set of Feynman diagrams where a scalar Higgs boson decays to two other Higgs through the triple Higgs coupling since this set of diagrams give a vanishing contribution, similar to class-A.

\begin{figure}[h]
	\centering
	\begin{subfigure}[c]{.2\textwidth}
		\raisebox{-\height}{\includegraphics[width=0.6\textwidth,angle=90
		]{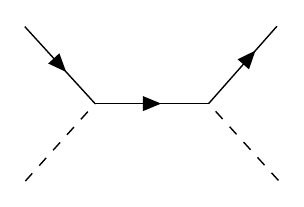}}
	\end{subfigure}
\hspace{1cm}
	\begin{subfigure}[c]{.2\textwidth}
		\raisebox{-\height}{\includegraphics[width=0.8\textwidth,angle=90]{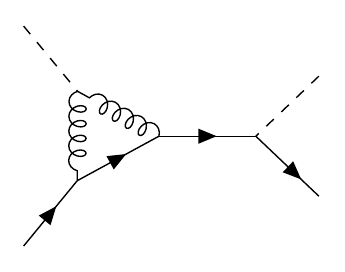}}
	\end{subfigure}
\hspace{1cm}
	\begin{subfigure}[c]{.2\textwidth}
		\raisebox{-\height}{\includegraphics[width=0.8\textwidth,angle=270]{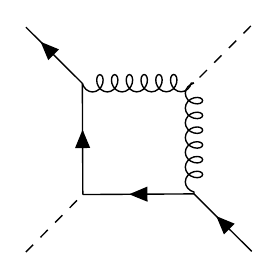}}
	\end{subfigure}
\hspace{1cm}
	\begin{subfigure}[c]{.2\textwidth}
		\raisebox{-\height}{\includegraphics[width=0.8\textwidth,angle=180]{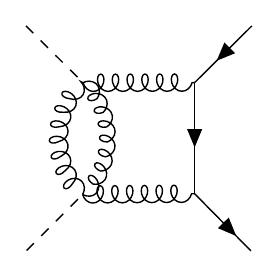}}
	\end{subfigure}
\hspace{1cm}
	\begin{subfigure}[c]{.2\textwidth}
		\raisebox{-\height}{\includegraphics[width=0.8\textwidth,angle=180]{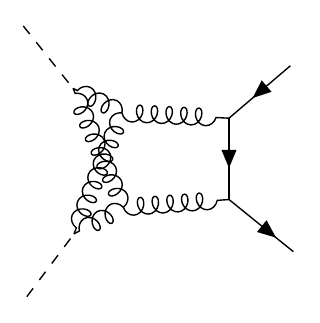}}
	\end{subfigure}
\hspace{1cm}
	\begin{subfigure}[c]{.2\textwidth}
		\raisebox{-\height}{\includegraphics[width=0.8\textwidth,angle=270]{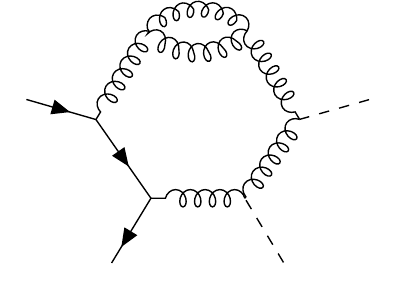}}
	\end{subfigure}
	\caption{Sample diagrams for $q\bar q\rightarrow\tilde\phi\tilde\phi$ at NNLO. The first diagram belongs to $O_JO_J$ category. The remaining two diagrams in the first row belong to $O_G O_J$. The diagrams in the second row fall under $O_GO_G$.}
	\label{fig:qqaa-classB}
\end{figure}

For the process $\tilde\phi\tilde\phi$ at LO, there are 2 diagrams, as shown in figure~\ref{fig:qqhh-classA}, containing the four-point vertex $gg\tilde\phi\tilde\phi$. These diagrams also vanish to all orders for the same reason as for the di-scalar. We categorise these as class-A. For higher orders, we classify the diagrams into three types. If two pseudo-scalars interact through only gluons: $O_G O_G$, only quarks: $O_J O_J$ and both gluon and quark: $O_GO_J$. At the NLO, which are of ${\cal O}(a_s^3)$, there can only be one-loop diagrams of class $O_GO_G$ similar to the first diagram in figure~\ref{fig:qqhh-classA}. There are 5 diagrams at this loop order. At the NNLO, we have all three types of diagrams, as shown in figure~\ref{fig:qqaa-classB}.

To apply the Feynman rules, perform SU($n_c$) color manipulation, and $d$-dimensional Lorentz as well as spinor algebra, we pass  the resulting expression through a series of in-house routines based on \form~\cite{Vermaseren:2000nd} and Mathematica. By employing \reduze~\cite{Studerus:2009ye,vonManteuffel:2012np}, we find the appropriate loop momentum shifts to recast each Feynman diagram beyond the tree level to a form belonging to one of the integral families. We have 3 and 6 integral families at one- and two-loop, respectively. The resulting expressions contain Feynman loop integrals which are reduced to a set of master integrals through integration-by-parts (IBP)~\cite{Chetyrkin:1981qh} identities. To perform the IBP reductions, we use \litered~\cite{Lee:2012cn,Lee:2013mka}\footnote{We thank Roman Lee for providing the yet to be published version of \litered.} in conjunction with \mint~\cite{Lee:2013hzt}. The latter assists to determine the number of master integrals. We get 10 and 149 master integrals at one- and two-loop, respectively. Using the results of the master integrals~\cite{Gehrmann:2013cxs,Gehrmann:2014bfa} in terms of multiple polylogarithms, we get the bare form factors as Laurent series in dimensional regulator $\epsilon$. To do the series expansion efficiently, we make use of the finite field reconstruction through \finiteflow{}~\cite{Peraro:2019svx}.

\section{Ultraviolet divergences and operator renormalisation}
\label{sec:op-ren}
Amplitude beyond LO often suffers from ultraviolet (UV) divergence. Since we are working in massless QCD, the first part of the cure consists of renormalising the strong coupling constant in $\overline{\rm MS}$ scheme through~\cite{Tarasov:1980au}
\begin{align} 
\label{eq:asren}
 \hat{a}_s\mu^{2\epsilon} S_{\epsilon} &= a_s(\mu_R)\mu_R^{2\epsilon} Z_a(\mu_R) \nonumber\\
 &= a_s\mu_R^{2\epsilon} \left[   1 - a_s \left( \frac{1}{\epsilon} \beta_0 \right) + {\cal O}(a_s^2)  \right]
\end{align}
where $S_{\epsilon} = {\rm exp} \left[(\ln 4\pi-\gamma_E)\epsilon
\right]$ with $\gamma_E \approx 0.5772$, the Euler-Mascheroni constant and $a_s(\mu_R) \equiv a_s$. The symbol hat ($~\hat{}~$) indicates the bare parameter, $Z_a$ is the coupling renormalisation constant, $\mu_0$ is a parameter in bare theory to keep the bare coupling dimensionless in dimensional regularisation, and $\beta_0=11/3 C_A-2/3 n_f$ is the leading coefficient of QCD $\beta$-function.
~\\

\textbf{Scalar Higgs}: Owing to working with composite operator, in addition to performing coupling constant renormalisation, we need to perform an additional operation to get rid of the full UV divergences. Since this is a property associated to the operator itself, we call it as operator renormalisation. For di-scalar, the renormalised Lagrangian, $[{\cal L}^\phi_{\rm eff}]_R$, is related to the bare part through
\begin{align}
\label{eq:ren-eff-Lag-phi}
[{\cal L}^\phi_{\rm eff}]_R = Z_\phi {\cal L}^\phi_{\rm eff}
\end{align}
with the operator renormalisation constant in ${\overline{\rm MS}}$ scheme~\cite{Nielsen:1975ph,Spiridonov:1988md,Kataev:1981gr}
\begin{align}
\label{eq:ope-ren-phi}
Z_\phi = 1-a_s \left(\frac{1}{\epsilon} \beta_0 \right) + {\cal O}(a_s^2)\,.
\end{align}
Incorporating the aforementioned renormalisations, we get the UV  renormalised FF
\begin{align}
\label{eq:ren-Fphi}
\left[F_\phi\right]_R = Z_\phi^2  F_\phi(a_s)
\end{align}
where $F_\phi$ appearing on the right-hand side is the one that we calculate by applying the projector \eqref{eq:projector} on the set of Feynman diagrams and performing coupling constant renormalisation through \eqref{eq:asren}. By rewriting $F_\phi$ in \eqref{eq:ren-Fphi} after factoring out the Wilson coefficient, we get 
\begin{align}
\label{eq:Fexpand-phi}
F_\phi = C_H^2(a_s) \sum_{n=1} \hat{a}_s^n \hat{F}^{(n)}_\phi\,,
\end{align}
where $\hat{F}^{(n)}_\phi$ corresponds to $n$-th loop bare form factor obtained by evaluating the Feynman diagrams in figure~\ref{fig:qqhh-classB}. By performing the required renormalisations and expanding all quantities in powers of $a_s$ as
\begin{align}
\label{eq:expand-Zphi-CH-symbolic}
&Z_\sigma = 1 + \sum_{j=1} a_s^j Z_\sigma^{(j)}, \quad C_H = \sum_{j=1} a_s^j C_H^{(j)}\,,
\end{align}
we arrive at the fully UV renormalised FF up to NNLO expanded in powers of renormalised $a_s$
\begin{align}
\label{eq:Fphi}
[F_\phi]_R &= a_s^3 \left(C_H^{(1)}\right)^2 \hat{F}^{(1)}_\phi + a_s^4 \left\{2 C_H^{(1)} C_H^{(2)} \hat{F}^{(1)}_\phi +  \left(C_H^{(1)}\right)^2 \left( \hat{F}^{(2)}_\phi +  \hat{F}^{(1)}_\phi Z_a^{(1)}  + 2 \hat{F}^{(1)}_\phi Z_\phi^{(1)}\right)\right\}\nonumber\\
&\equiv a_s^2 \sum_{k=1}^2 a_s^k [F_\phi^{(k)}]_R\,.
\end{align}
In the aforementioned equation, the renormalised FFs at NLO and NNLO are denoted by $[F_\phi^{(1)}]_R$ and $[F_\phi^{(2)}]_R$, respectively. We provide the corresponding finite remainders in the ancillary file with the \arXiv~submission. The coefficients $Z_\sigma^{(j)}$ and $C_H^{(j)}$ in \eqref{eq:expand-Zphi-CH-symbolic} can be extracted from their explicit results in \eqref{eq:asren}, \eqref{eq:ope-ren-phi} and \eqref{eq:CH}, where $\sigma$ stands for $a$ and $\phi$. Before moving to the next topic, we point out that unlike the case of di-scalar Higgs boson production through gluon fusion~\cite{Zoller:2016iam,Banerjee:2018lfq}, we do not need any additional renormalisation arising from the contact terms because of the product of two composite operators at a short distance. Although, in principle, one could expect it to be present.
~\\

\textbf{Pseudo-scalar Higgs}: In the case of di-pseudo scalar Higgs boson, we also have to perform coupling constant and operator renormalisation in $\overline{\rm MS}$ scheme, for example. However, for the operator $O_J$, which is constructed out of the singlet axial vector current $\bar\psi \gamma^\mu\gamma_5\psi$, there is another phenomenon takes place due to the fact that our scheme of chiral quantity depicted through \eqref{eq:gamma5-defn} violates the defining anti-commuting property of $\gamma_5$ in the evanescent $4-d=2\epsilon$ dimensional space. Consequently, this gives rise to additional terms of UV-divergence origin in the dimensionally regularised amplitudes which need to be removed~\cite{Chanowitz:1979zu,Trueman:1979en} manually. This is manifested as a loss of correct chiral Ward identity. To cure this phenomenon, we need to introduce additional UV renormalisation constants~\cite{Trueman:1979en,Larin:1991tj,Larin:1993tq,Kodaira:1979pa} in addition to the pure $\overline{\rm MS}$ renormalisation constant. Therefore, the renormalised operator $[O_J]_R$ can be written as
\begin{align}
\label{eq:OJ-ren}
[O_J]_R = Z^s_5 Z^s_{\overline{\rm MS}} O_J \equiv Z_J O_J
\end{align}
where the corresponding bare quantity $O_J$ is defined in \eqref{eq:OGOJ}. We denote the total operator renormalisation constant of the axial vector current by $Z_J$ consisting of a pure $\overline{\rm MS}$ renormalisation factor $Z^s_{\overline{\rm MS}}$ and an additional finite renormalisation part $Z^s_5$. The latter is needed to restore the correct form of the axial Ward identity which is known to be anomalous. With the recent result of the $Z^s_5$ by one of the authors of this article, it is now available to ${\cal O}(a_s^3)$~\cite{Larin:1993tq,Ahmed:2021spj}. 
The $Z^s_{\overline{\rm MS}}$ is available to ${\cal O}(a_s^3)$ in refs.~\cite{Larin:1993tq,Ahmed:2015qpa,Ahmed:2021spj}.

On the other hand, unlike the previous case, the axial anomaly operator $O_G$ is not renormalised multiplicatively~\cite{Adler:1969gk} but rather undergoes mixing with the divergence of the axial current operator, $O_J$ as~\cite{Espriu:1982bw,Breitenlohner:1983pi}
\begin{align}
\label{eq:OG-ren}
    [O_G]_R = Z_{G\tilde G} O_G + Z_{GJ} O_J\,.
\end{align}
The $Z_{G\tilde G}$ is available to ${\cal O}(a_s^4)$ in refs.~\cite{Espriu:1982bw,Larin:1993tq,Bos:1992nd,Ahmed:2015qpa,Ahmed:2021spj} which coincides with the strong coupling constant renormalisation $Z_a$, a result  claimed to hold to all orders in ref.~\cite{Breitenlohner:1983pi} and proved explicitly in ref.~\cite{Luscher:2021bog}. The other constant $Z_{GJ}$ which starts at ${\cal O}(a_s)$ is computed to ${\cal O}(a_s^3)$ in refs.~\cite{Larin:1993tq,Zoller:2013ixa,Ahmed:2015qpa}.

For the di-scalar production, effectively, we have to renormalise three different combinations of operators corresponding to three categories of Feynman diagrams which are
\begin{align}
\label{eq:ren-tildephi-operators}
&[O_GO_G]_R = Z_{G\tilde G}^2 O_G O_G + 2 Z_{G\tilde G} Z_{GJ} O_G O_J + Z_{GJ}^2 O_J O_J\,,\nonumber\\
&[O_GO_J]_R = Z_{G\tilde G} Z_{J} O_G O_J +  Z_{GJ} Z_{J} O_J O_J\,,\nonumber\\
&[O_JO_J]_R = Z_{J}^2 O_J O_J\,.
\end{align}
Inserting these into external quark states, we arrive at the following form of the UV renormalised FFs:
\begin{align}
\label{eq:ren-FF-tildephi}
[F_{\tilde \phi}]_R = [F_{\tilde\phi,G G}]_R + [F_{\tilde\phi,GJ}]_R + [F_{\tilde\phi,JJ}]_R\,,
\end{align}
where these individual components are given by
\begin{align}
\label{eq:ren-FF-tildephi-comp}
&[F_{\tilde\phi,G\tilde G}]_R = C_G^2 \left(Z_{G\tilde G}^2 F_{\tilde\phi,G G} (a_s) + 2 Z_{G\tilde G} Z_{GJ} F_{\tilde\phi,GJ} (a_s) + Z_{GJ}^2 F_{\tilde\phi,JJ} (a_s) \right)\,,\nonumber\\
&[F_{\tilde\phi,GJ}]_R = C_G C_J \left(Z_{G\tilde G}Z_J F_{\tilde\phi,GJ} (a_s) +  Z_{GJ} Z_J F_{\tilde\phi,JJ} (a_s) \right)\,,\nonumber\\
&[F_{\tilde\phi,JJ}]_R = C_J^2 Z_J^2 F_{\tilde\phi,JJ}\,.
\end{align}
The FF, $F_{\tilde\phi,G_1G_2}$, appearing on the right-hand side of the above equations are obtained by applying the projector \eqref{eq:projector} on the set of Feynman diagrams and performing coupling constant renormalisation through \eqref{eq:asren}. The $\{G_1G_2\}$ can be $\{GG,GJ,JJ\}$ depending on which set of diagrams we are considering, as shown in figure~\ref{fig:qqaa-classB}. While writing down the \eqref{eq:ren-FF-tildephi-comp}, we explicitly factor out the Wilson coefficients. By expressing the form factors in terms of bare quantities at $n$-th loop, $\hat{F}^{(n)}_{\tilde{\phi},G_1G_2}$, as
\begin{align}
\label{eq:Fexpand-phitilde-gg}
&F_{\tilde\phi,GG} = \sum_{n=1} \hat{a}_s^n \hat{F}_{\tilde\phi,GG}^{(n)}\,,\nonumber\\
&F_{\tilde\phi,GJ} =  \sum_{n=1}  \hat{a}_s^n \hat{F}_{\tilde\phi,GJ}^{(n)}\,,\nonumber\\
&F_{\tilde\phi,JJ} =  \sum_{n=0}  \hat{a}_s^n \hat{F}_{\tilde\phi,JJ}^{(n)}\,.
\end{align}
The bare FF are the ones which are obtained by applying the projector directly on the set of Feynman diagrams in figures~\ref{fig:qqaa-classB}. Note that the leading powers of the strong coupling constant in \eqref{eq:Fexpand-phitilde-gg} are determined excluding the Wilson coefficients since the latter are factored out in \eqref{eq:ren-FF-tildephi-comp}.
For instance, $\hat{F}_{\tilde\phi,GJ}^{(n)}$ corresponds to the diagrams involving $O_GO_J$. To reach at a more transparent form of the complete UV renormalised FF, we expand the operator renormalisation constants, introduced in \eqref{eq:OJ-ren} and \eqref{eq:OG-ren}, and the Wilson coefficients in powers of the renormalised coupling constant $a_s$ as
\begin{align}
\label{eq:expand-Zij-symbolic}
&Z_{G\tilde G} = 1 + \sum_{j=1} a_s^j Z_{G\tilde G}^{(j)}, \quad Z_{GJ} = a_s Z_{GJ}^{(1)}, \nonumber\\
&Z_J =\Big(1+\sum_{j=1} a_s^j Z_5^{s,(j)} \Big) \Big(1 + \sum_{k=2} a_s^k Z_{\overline{\rm MS}}^{s,(k)} \Big)\,,\nonumber\\
&C_G = a_s C_G^{(1)}\,,\quad C_J = \sum_{j=2} a_s^j C_J^{(j)}\,.
\end{align}
The notations are exactly similar to the ones in \eqref{eq:expand-Zphi-CH-symbolic}. The lower limit in each of these expansions is written as per their leading order expressions. By incorporating these, we arrive at the following fully UV renormalised FF, introduced in  \eqref{eq:ren-FF-tildephi} and \eqref{eq:ren-FF-tildephi-comp}:
\begin{align}
\label{eq:Fphitilde-Renorm}
[F_{\tilde\phi,G\tilde G}]_R &= a_s^3 \left(C_G^{(1)}\right)^2 \hat{F}^{(1)}_{\tilde\phi,GG} \nonumber\\&+ a_s^4   \left(C_G^{(1)}\right)^2 \Big( \hat{F}^{(2)}_{\tilde\phi,GG} + Z_a^{(1)} \hat{F}^{(1)}_{\tilde\phi,GG} + 2 Z_{G\tilde G}^{(1)} \hat{F}^{(1)}_{\tilde\phi,GG} + 2 Z_{GJ}^{(1)} \hat{F}^{(1)}_{\tilde\phi,GJ} + \left(Z_{GJ}^{(1)}\right)^2 \hat{F}^{(0)}_{\tilde\phi,JJ} \Big)  \,,\nonumber\\
[F_{\tilde\phi,GJ}]_R &= a_s^4 C_G^{(1)} C_J^{(2)} \left( \hat{F}^{(1)}_{\tilde\phi,GJ} + Z_{GJ}^{(1)} \hat{F}^{(0)}_{\tilde\phi,JJ} \right) \,, \nonumber\\
[F_{\tilde\phi,JJ}]_R &= a_s^4 \left( C_J^{(2)} \right)^2 \hat{F}^{(0)}_{\tilde\phi,JJ} \,.
\end{align}
By adding these three contributions order-by-order, we get the final UV renormalised FF at NLO and NNLO, $[F_{\tilde \phi}^{(1)}]_R$ and $[F_{\tilde \phi}^{(2)}]_R$, respectively, with
\begin{align}
\label{eq:FF-Ren-tildephi-expansion}
[F_{\tilde \phi}]_R = a_s^2 \sum_{k=1}^2 a_s^k [F_{\tilde \phi}^{(k)}]_R\,.
\end{align}
We provide the finite remainder of FF at NLO and NNLO as an ancillary file with the \arXiv~submission. Interestingly, similar to the case of di-scalar Higgs boson, we find that we do not need any additional renormalisation arising from the contact terms owing to the product of operators at a short distance. For the gluonic channel, its absence is shown both from the operator product expansion~\cite{Zoller:2013ixa} and direct two-loop calculation~\cite{Bhattacharya:2019oun}. In the next section, we talk about the infrared structure of the UV renormalised FF and subsequently the behaviour of finite remainders.

\section{Infrared factorisation and finite remainders}
\label{sec:IR}
The UV renormalised FF still contains divergences originated from the soft and collinear configurations of loop momentum, which can be parameterised in terms of a universal subtraction operator $\textbf{I}^{(1)}(\epsilon)$ as~\cite{Catani:1998bh,Sterman:2002qn}
\begin{align}
\label{eq:IR-factorisation}
[F^{(2)}_h]_R = 2 \textbf{I}^{(1)}(\epsilon) [F^{(1)}_h]_R + [F^{(2)}_{h,{\rm fin}}]_R
\end{align}
where $[F^{(2)}_{h,{\rm fin}}]_R$ at $\epsilon \rightarrow 0$ is the finite remainder. The IR subtraction operator for the processes under consideration in conventional dimensional regularisation (CDR) is given by
\begin{align}
\label{eq:I1}
\textbf{I}^{(1)}(\epsilon) = -C_F \frac{e^{\epsilon\gamma_E}}{\Gamma(1-\epsilon)} \left( \frac{1}{\epsilon^2} + \frac{3}{2\epsilon} \right) \left( \frac{-\mu_R^2}{s} \right)^\epsilon\,.
\end{align}
We check that the renormalised FF do exhibit this predicted infrared structure. The double pole is checked analytically, whereas the single one is done numerically due to the complexity of the algebraic expression. We check it on several set of kinematic points. This serves as a very strong check on the correctness of our computation. By subtracting the IR singular terms from the UV renormalised FF, we obtain the finite remainders for all the FF. These finite remainders will contribute to observables, say the inclusive cross-section, at the NNLO for the process of di-Higgs or di-pseudo-Higgs boson productions through quark annihilation at hadron collider. 
In the upcoming subsections, we discuss the kinematic symmetry associated with the interchange of two final state particles and the behaviour of leading transcendental weight terms.

\subsection{Permutation symmetry}
\label{ss:symm}
The UV renormalised FF or equivalently the finite remainder of the FF for both processes should exhibit the kinematic symmetry associated with the interchange of two final state particles. Due to the presence of an initial state quark and anti-quark pair, the FF are expected to be antisymmetric upon interchanging either the momenta of two incoming ($p_1 \leftrightarrow p_2$) or two outgoing ($p_3 \leftrightarrow p_4$) particles. This gets translated to an interchange of the Mandelstam variables, $t \leftrightarrow u$, or the scaleless parameters, $y \leftrightarrow z$, as introduced in \eqref{eq:xyz}. This implies 
\begin{align}
\label{eq:kinematic-symm-FF}
[F_h^{(k)}(y,z)]_R + [F_h^{(k)}(z,y)]_R = 0 \,.
\end{align}
We check that the resulting FF for both the processes do exhibit the expected kinematic symmetry by verifying the above equation at one and two-loop. Although this checking is quite simple for one-loop, this is quite involved at two-loop owing to the presence of special functions such as $\rm{Li_{2,2}}$. With the help of a C++ code~\cite{Frellesvig:2016ske}, we evaluate this special function numerically and verify the kinematic symmetry. This serves as a strong check on the finite part of the FF.

\subsection{Behaviour of leading transcendentality weight terms}
\label{sec:UT}
Understanding the connection among various gauge theories through studying the analytic structures of scattering amplitudes is an active area of investigation. In particular, the connection between maximally supersymmetric Yang-Mills theory (mSYM) and QCD is of fundamental importance. In refs.~\cite{Kotikov:2001sc,Kotikov:2004er,Kotikov:2006ts,Banerjee:2018yrn}, the anomalous dimensions of leading twist-two-operator in mSYM are found to be identical to the highest transcendental (HT) weight counterparts in QCD, and consequently, the principle of maximal transcendentality (PMT) is conjectured. This principle, albeit an observational fact, states that algebraically the HT weight terms of certain quantities in mSYM and QCD are identical upon changing the representation of fermions in QCD from fundamental to adjoint through $C_A=C_F=2 n_f T_F=n_c$ with $T_F$ being the normalisation factor in fundamental representation. The HT weight terms of quark and gluon two-point FF with a single operator insertion in QCD are found to be identical to the scalar FF of half-BPS primary in mSYM to three loops~\cite{Gehrmann:2011xn} (up to some normalisation factor of $2^L$ with $L$ being the loop).  The diagonal terms of the two-point pseudo-scalar~\cite{Ahmed:2015qpa} and stress energy tensorial FF~\cite{Ahmed:2015qia,Ahmed:2016qjf} also obey the conjecture. Moreover, the three-point scalar and pseudo-scalar FF are also found to respect the PMT~\cite{Brandhuber:2012vm,Loebbert:2015ova,Loebbert:2016xkw,Banerjee:2017faz,Brandhuber:2017bkg,Jin:2018fak,Brandhuber:2018xzk,Jin:2019ile,Jin:2019opr}. In ref.~\cite{Dixon:2017nat}, the four loop collinear anomalous dimension in the planar mSYM is computed employing this conjecture. Beyond form factors, even some observables, such as the asymptotic limit of energy-energy correlator~\cite{Belitsky:2013ofa} and soft function~\cite{Banerjee:2018yrn} are also observed to be consistent with PMT. 

As the domain of validity of this conjecture is still not well established, the natural question arises whether the PMT carries over to more general class of FF and correlation functions. Already some instances of the violation of this conjecture are found. The three-point quark~\cite{Ahmed:2016yox} and gluon~\cite{Ahmed:2014gla} FF with a single stress energy tensorial operator insertion~\cite{Ahmed:2019nkj}, and Regge limit of amplitudes~\cite{DelDuca:2017peo} do not obey the principle. In refs.~\cite{Ahmed:2019upm,Ahmed:2019yjt}, it has been shown that the conjecture also fails to hold true for two-point FF with two operators insertion. In view of this, we investigate whether there is any similarity between the HT weight terms of the scalar and pseudo-scalar FF as computed in this article and that of mSYM theory. We find they do not share the same HT weight terms at one- as well as two-loop. For instance, the one-loop finite remainders contain three weight 2 polylogarithms, ${\rm Li_2}(-x)$, ${\rm Li_2}(1/(1+y))$ and ${\rm Li_2}(1/(1+z))$, with the coefficients
\begin{align}
&{\rm Cof}_{{\rm Li_2}(-x)} = -i  w^2_{h} C_F ~ \frac{(1 + x^4) (1 + x^2 - 2 x y)}{x (x^2-1)(y + x^2 y - x (1 + y^2))} \,,\nonumber\\
&{\rm Cof}_{{\rm Li_2}(1/(1+y))} =   i  w^2_{h} C_F ~\frac{  y + x^2 y - x (1 + 2y^2) }{y + x^2 y - x (1 + y^2)} \,,\nonumber\\
&{\rm Cof}_{{\rm Li_2}(1/(1+z))} =  i w^2_{h} C_F ~\frac{1 + x^4 - 3 x y - 3 x^3 y + x^2 (3 + 2y^2)}{x (y + x^2 y - x (1 + y^2))}
\end{align}
where $w_{\phi} = -4/3$ for the scalar Higgs boson and $w_{\tilde \phi} = -2^{\frac{5}{4}} G_{F}^{\frac{1}{2}} {\rm \cot} \beta$ for the pseudo-scalar Higgs boson production. The corresponding one-loop two-point scalar FF in mSYM~\cite{Ahmed:2019yjt} with two operator insertion of half-BPS primary contain different kinematic coefficients.

\begin{figure}[h!]
\centering
\includegraphics[scale=0.5]{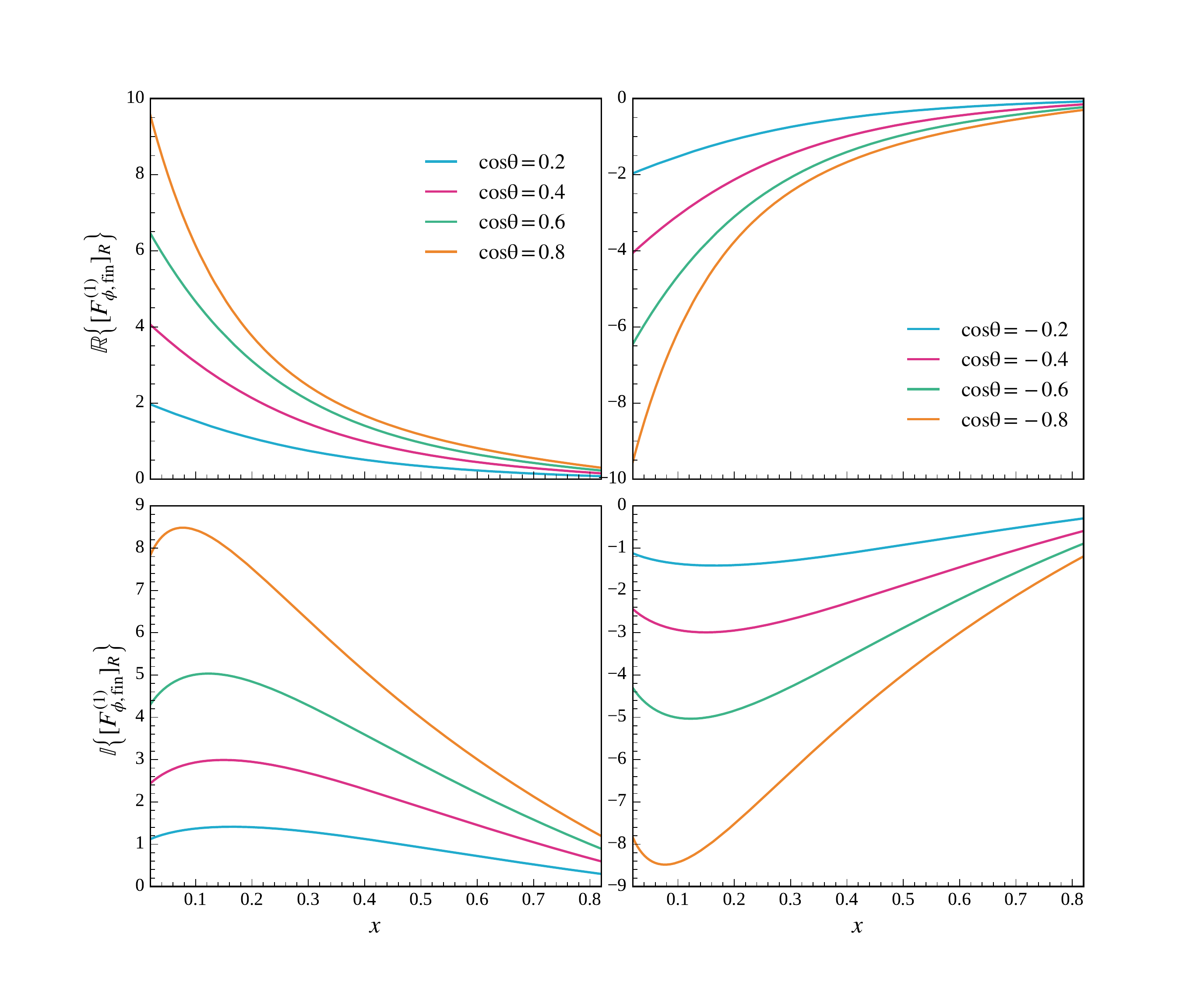}
\caption{Behaviour of real (upper panel) and imaginary parts (lower panel) of one-loop UV renormalised finite remainder \Big($\mathbb{R} \left\{ [F^{(1)}_{\phi,{\rm fin}}]_R \right\}$ and $\mathbb{I} \left\{ [F^{(1)}_{\phi,{\rm fin}}]_R \right\}$\Big) for the process $q+\bar q \rightarrow \phi + \phi$ as a function of the scaling variable $x$ for positive (left panel) and negative (right panel) values of $\cos(\theta)$.}
\label{fig:oneL}
\end{figure} 
\section{Numerical analysis of the results} \label{sec:Num}
In this section, we study the behaviour of the finite remainders $[F^{(1)}_{h,{\rm fin}}]_R $ at one loop for the production of di-Higgs and di-pseudo-Higgs boson through quark annihilation in the large top quark mass limit. The results are expressed in terms of the classical polylogarithms with the scaling variables $x$ and $y$ being their arguments. In figure \ref{fig:oneL}, we plot the real (upper panel) and imaginary (lower panel) parts of $[F^{(1)}_{h,{\rm fin}}]_R $ as a function of the variable $x$ for different positive (left panel) and negative (right panel) values of $\cos(\theta)$, where $\theta$ is the angle between one of the (pseudo-) Higgs bosons in the final state and one of the initial quarks in their centre of mass frame. Note that the finite remainders $[F^{(1)}_{h,{\rm fin}}]_R $ at one loop for  di-Higgs and di-pseudo-Higgs bosons differ only by their respective Wilson coefficients, hence we provide only one plot which is for the production of di-scalar Higgs boson. It is evident from the plots that the amplitudes are anti-symmetric under $\cos(\theta) \rightarrow - \cos(\theta)$, as expected for a purely fermionic amplitude. As a result, the amplitude vanishes when $\cos(\theta)=0$.
Since this anti-symmetry property has not been used in our calculation, it, in fact, serves as a strong check on our results. We observe that the numerical values of real as well as imaginary parts of $[F^{(1)}_{h,{\rm fin}}]_R $ corresponding to different values of $\cos(\theta)$ start to converge as $x \rightarrow 1$.

\section{Conclusions and Outlook}
\label{sec:concl}
In this article, we present the two-loop four-point amplitudes for the productions of two scalar and two pseudo-scalar Higgs bosons through quark annihilation in Higgs effective field theory. The HEFT is obtained by treating the top quark as infinitely heavy. We perform the calculation under dimensional regularisation. The chiral quantities, Dirac's $\gamma_5$ and $\varepsilon_{\mu\nu\rho\sigma}$, are treated through Larin's prescription. The method of projection is employed to perform the calculation. From the results, one can extract the helicity amplitudes in 't Hooft-Veltman scheme. The ultraviolet renormalisation includes the strong coupling constant as well as operator renormalisations. For the pseudo-scalar, the latter involves operator mixing. 

In any Greeen's function containing multiple local operators, one could expect the presence of an additional divergence that could arise from the short distance behaviour of the two operators. However, we find no additional divergence of this kind in both the amplitudes. Consequently, there is no contact renormalisation that is needed. This is in corroboration with the analysis through operator product expansion.

The fully ultraviolet renormalised amplitudes exhibit the infrared poles in dimensional regulator as dictated by the prediction in terms of Catani's universal subtraction operator. We check that the finite remainders exhibit expected kinematic symmetry. We also implement the finite remainders in a numerical code that can be used for further phenomenological studies. The finite remainders are provided as ancillary files. Given the availability of these results, we are now one step closer to make the prediction of an observable for the production of di-Higgs or di-pseudo-Higgs arising from a subdominant channel which is important at this time of very high precision physics.

\section*{Acknowledgements}
The work of T.A. received funding from the European
Research Council (ERC) under the European Union’s Horizon 2020 research and innovation programme \textit{High precision multi-jet dynamics at the LHC} (ERC Condsolidator grant agreement No 772009). TA thanks S. Zoia for helping in series expansion employing the technique of finite field reconstruction. TA also gratefully acknowledges the computing resources
provided by Max-Planck-Institute for Physics.

\appendix
\section{Results of the one-loop form factors}
\label{secA:1-loop}
In this appendix, we present the finite remainders, as defined in \eqref{eq:IR-factorisation}, of the form factors at one loop for both scalar and pseudo-scalar Higgs boson productions through quark annihilation. We present the real, $\mathbb{R} \left\{ [F^{(1)}_{h,{\rm fin}}]_R \right\}$ and imaginary parts, $\mathbb{I} \left\{ [F^{(1)}_{h,{\rm fin}}]_R \right\}$ separately which are
\begin{align}
\begin{autobreak}
    \mathbb{R} \left\{ [F^{(1)}_{h,{\rm fin}}]_R \right\}=

       w^2_h C_F \pi {\rm Den}(x,y) \Bigg[ \log(1+z)    \bigg\{
          4
          - 12 x y
          + 8 x^2
          + 8 x^2 y^2
          - 8 x^4
          - 8 x^4 y^2
          + 12 x^5 y
          - 4 x^6            \bigg\}

       +\log(1+y)   \bigg\{
           4 x y
          - 4 x^2
          - 8 x^2 y^2
          + 4 x^4
          + 8 x^4 y^2
          - 4 x^5 y
           \bigg\}

       +\log(1+x)    \bigg\{
          - 4
          + 8 x y
          - 4 x^2
          + 4 x^4
          - 8 x^5 y
          + 4 x^6
           \bigg\}

       +\log(z)  \bigg\{
          - 2
          + 6 x y
          - 4 x^2
          - 4 x^2 y^2
          + 4 x^4
          + 4 x^4 y^2
          - 6 x^5 y
          + 2 x^6
           \bigg\}

       +\log(x)    \bigg\{
           2
          - 4 x y
          + 2 x^2
          - 2 x^4
          + 4 x^5 y
          - 2 x^6
           \bigg\}

       +\log(y)   \bigg\{
          - 2 x y
          + 2 x^2
          + 4 x^2 y^2
          - 2 x^4
          - 4 x^4 y^2
          + 2 x^5 y
           \bigg\} \Bigg] \,,

\end{autobreak}    \\
\begin{autobreak}
\mathbb{I} \left\{ [F^{(1)}_{h,{\rm fin}}]_R \right\} = 

           w^2_{h} C_F {\rm Den}(x,y) \Bigg[\bigg\{
          - 4 x^4 \zeta_2
          + 8 x^5 \zeta_2 y
          - 4 x^6 \zeta_2
           \bigg\}

       + {\rm{Li}}_2(-x)    \bigg\{
          - 4
          + 8 x y
          - 4 x^2
          - 4 x^4
          + 8 x^5 y
          - 4 x^6
           \bigg\}

       +{\rm{Li}}_2(1/(1 + z))    \bigg\{
          - 4
          + 12 x y
          - 8 x^2
          - 8 x^2 y^2
          + 8 x^4
          + 8 x^4 y^2
          - 12 x^5 y
          + 4 x^6
           \bigg\}

       +{\rm{Li}}_2(1/(1+y))   \bigg\{
          - 4 x y
          + 4 x^2
          + 8 x^2 y^2
          - 4 x^4
          - 8 x^4 y^2
          + 4 x^5 y
           \bigg\}

       + \log^2(1 + z)   \bigg\{ - 2
          + 6 x y
          - 4 x^2
          - 4 x^2 y^2
          + 4 x^4
          + 4 x^4 y^2
          - 6 x^5 y
          + 2 x^6 \bigg\}

       + \log^2(1+y)  \bigg\{- 2 x y
          + 2 x^2
          + 4 x^2 y^2
          - 2 x^4
          - 4 x^4 y^2
          + 2 x^5 y
           \bigg\}

       + \log(1+x) \log(z)     \bigg\{ 4
          - 12 x y
          + 8 x^2
          + 8 x^2 y^2
          - 8 x^4
          - 8 x^4 y^2
          + 12 x^5 y
          - 4 x^6 \bigg\}

       + \log(z)    \bigg\{
          - 3 x y
          + 3 x^2
          + 3 x^2 y^2
          - 3 x^4
          - 3 x^4 y^2
          + 3 x^5 y
           \bigg\}

       + \log^2(z)    \bigg\{
           1
          - 3 x y
          + 2 x^2
          + 2 x^2 y^2
          - 2 x^4
          - 2 x^4 y^2
          + 3 x^5 y
          - x^6
           \bigg\}

       + \log(x) \log(z)    \bigg\{
          - 2
          + 6 x y
          - 4 x^2
          - 4 x^2 y^2
          + 4 x^4
          + 4 x^4 y^2
          - 6 x^5 y
          + 2 x^6
           \bigg\}

        + \log(x)^2    \bigg\{
          - 1
          + 2 x y
          - x^2
          - x^4
          + 2 x^5 y
          - x^6
           \bigg\}

       + \log(x) \log(y)     \bigg\{
          - 2 x y
          + 2 x^2
          + 4 x^2 y^2
          - 2 x^4
          - 4 x^4 y^2
          + 2 x^5 y
           \bigg\}

       + \log(y)     \bigg\{
           3 x y
          - 3 x^2
          - 3 x^2 y^2
          + 3 x^4
          + 3 x^4 y^2
          - 3 x^5 y
           \bigg\}

       + \log(y) \log(1+x)     \bigg\{
           4 x y
          - 4 x^2
          - 8 x^2 y^2
          + 4 x^4
          + 8 x^4 y^2
          - 4 x^5 y
           \bigg\}

       + \log^2(y)     \bigg\{   x y - x^2
          - 2 x^2 y^2
          + x^4
          + 2 x^4 y^2
          - x^5 y \bigg\} \Bigg] \,,
\end{autobreak}
\end{align}
where $w_{\phi} = -4/3$ for the scalar Higgs boson and $w_{\tilde \phi} = -2^{\frac{5}{4}} G_{F}^{\frac{1}{2}} {\rm \cot} \beta$ for the pseudo-scalar Higgs boson production. In the above expression, ${\rm Den}(x,y)$ is given by
\begin{align}
{\rm Den}(x,y) \equiv \frac{1}{4x (x^2-1) (x-y) (x y-1)} \,.    
\end{align}

\bibliography{qqhh} 
\bibliographystyle{utphysM}
\end{document}